\documentclass{ws-procs10x7}

\begin{document}

\title{Molecular Interpretation of the X(3872)}

\author{E.S. Swanson}

\address{Department of Physics and Astronomy\\ University of Pittsburgh\\ Pittsburgh PA 15260 USA
\\E-mail: swansone@pitt.edu}

\twocolumn[\maketitle\abstract{
The discovery of the $X(3872)$ in the $3\pi J/\psi$ mode is compelling evidence for 
its molecular nature. A successful prediction of this decay mode and other predictions
are reviewed here.
}]

In September of 2003 the Belle Collaboration announced the discovery of a narrow resonance in $\pi\pi J/\psi$ at 3872 MeV\cite{B}. The proximity of the state to $D^{0*} D^0$ threshold
generated immediate speculation that it is a $\bar D D^*$ molecular state\cite{CP,DD}.
This idea was developed into a predictive model in Ref.\cite{ess} by postulating that
the short and intermediate range dynamics which drive the structure of meson molecular
states are dominated by one pion and quark exchange processes. Thus the interaction Lagrangian was taken to be

\begin{eqnarray}
{\cal L} &=&  {1\over 2} \int d^3x d^3y \psi^\dagger(y)T^a\psi(y)\, 
K(x-y) \cdot  \\
&& \cdot \psi^\dagger(x) T^a \psi(x) + \\
&& {g\over \sqrt{2} f_\pi} \int d^4x \bar\psi(x) \gamma^\mu \gamma_5 \tau^a \psi(x) \cdot \\
&& \cdot \partial_\mu \pi^a(x).
\label{piq}
\end{eqnarray}
Here $f_\pi = 92$ MeV is the pion decay constant, $\tau$ is an SU(2) flavour generator, $g$ is a coupling determined by properties of deuteron, and $K$ is a kernel describing the interaction of constituent quarks. The parameters of the kernel are fixed by meson and baryon
spectroscopy. Finally a cut off is introduced to regulate the pion exchange potential and
separate short and long range interactions which is also fixed by the deuteron. Thus there are no 
adjustable parameters when
modelling meson molecular states.

One proceeds by projecting the quark-pion level interactions onto the channels of interest. These are $\bar D^0D^{0*}$ and  $D^+D^{-*}$ in S- and D-waves; there are also  two nearby 
hidden charm channels, $\rho J/\psi$ and $\omega J/\psi$. Although these channels are expected to be weak in the $X(3872)$ they are of central importance because they permit 
strong decay modes of the $X$ through the intrinsic widths of the $\rho$ and $\omega$.

The detailed computations of Ref. \cite{ess} reveal that only one bound state with 
$J^{PC} = 1^{++}$ exists. This state does not have good isospin because its binding 
energy is of order the mass splitting between the two charge modes of the $\bar D D^*$
channels. This is a generic feature of weakly bound molecular states and provides an 
important glimpse into any putative molecular state's structure. Since the prediction
of these quantum numbers, many additional measurements of $X(3872)$ properties have 
been made, and many possible $J^{PC}$ ({\it except} $1^{++}$) have been eliminated\cite{Olsen}.

Isospin symmetry breaking allows a weak $\rho J/\psi$ component of the $X(3872)$ which, through $\rho \to \pi\pi$, drives the $\pi\pi J/\psi$ discovery mode of the $X(3872)$. Similarly, the $\omega J/\psi$ component drives the $3\pi J/\psi$ decay mode. For weak 
binding the predicted widths  of Ref. \cite{ess} are 1.3 MeV for $2\pi J/\psi$ and $0.7$ MeV for $3\pi J/\psi$, yielding a branching fraction ratio of approximately 56\%. Furthermore,
the $3\pi$ Dalitz plot should have events near the edge of phase space, consistent with
production via a virtual $\omega$.
Note that
the $2\pi J/\psi$ prediction implies that the $2\pi \gamma J/\psi$ partial width is roughly
13 keV.

At ICHEP, the Belle collaboration announced that it had discovered the $X(3872)$ in the
predicted $3\pi J/\psi$ decay mode\cite{3pi}. The shape of the $3\pi$ spectrum is as expected 
and the measured $3\pi$ to $2\pi$ ratio is

\begin{equation}
{\Gamma(X \to \pi^+\pi^-\pi^0 J/\psi)\over \Gamma(X\to \pi^+\pi^-J/\psi)} = 0.8 \pm 0.3 \pm 0.1
\end{equation}
in remarkably good agreement with the prediction.

The only other mention of the $3\pi J/\psi$ decay mode in the literature comes from 
Ref. \cite{CP} where it is said that this mode is ``negligible".   
This statement has its roots in an assumed
pure $\bar D^0 D^{0*}$ content of the $X(3872)$. Thus decay to $3\pi J/\psi$ is driven
by virtual mixing to $\omega J/\psi$ via the line width of the $\omega$. However, the narrow
width of the $\omega$ (8 MeV) versus the large mass splitting between the $X$ and the $\omega J/\psi$ channel (7 MeV) implies that this mode should be small. This argument is
obviated in the current model by including the $\omega J/\psi$ channel in the bound state Schr\"odinger
equation. The result may be interpreted as an off-shell $\omega$ component of the $X$
which can subsequently decay via its full width.

Once a specific model of the $X$ has been adopted many predictions are possible. For example, radiative decays may be simply computed in the impulse approximation, as indicated in
figures 1 and 2. Predicted rates for $\gamma \psi'$, $\gamma \psi''$, and $\gamma \psi_2$ are very small in contrast to $c\bar c$ expectations of 10-100 keV. The predicted rate
for molecular $X$ to $\gamma J/\psi$ is 8 keV, at the low end of quark model
calculations of 10-140 keV for the $\chi_1'$ $c\bar c$ state.

\begin{figure}
\epsfxsize120pt
\figurebox{120pt}{160pt}{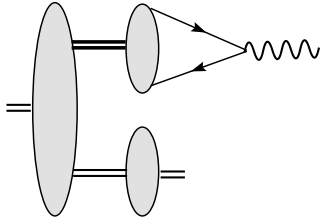}
\caption{Vector Meson Dominance Diagram.}
\label{VMD}
\end{figure}

\begin{figure}[h]
\includegraphics[angle=0,width=5cm]{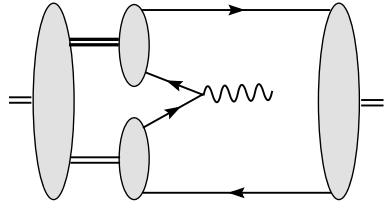}
\caption{\label{ANN} Annihilation Diagram}
\end{figure}

Finally, the rate to $\gamma\gamma$ is expected to be very small (much smaller than
typical excited $\chi$ states) because it either proceeds via a double vector meson
dominance diagram or an annihilation diagram where the photons are produced from
quarks in different mesons. Both diagrams are suppressed by a wavefunction at the 
origin which goes like $|u(0)|^2 \sim \sqrt{\mu_{DD^*}E_B}$
for weakly bound states.

The observation of the $X(3872)$ in $2\pi J/\psi$ and $3\pi J/\psi$ modes
implies that this state does not have good G-parity and hence does not have good isospin.
Clearly
the $X$ is unusual and it is a leading candidate for a molecular state.  Future
experimental determinations of its properties will settle the question.

\section*{Acknowledgments}

This work was supported by the DOE under contract DE-FG02-00ER41135.

\end{document}